# On the Optimality of Decode and Forward for Some Cooperative Broadcast Channels

Nicolas Le Gouic  
*LTCI Telecom Paris*  
91120 Palaiseau, France  
nicolas.legouic@ip-paris.fr

Yossef Steinberg  
*Technion-Israel Institute of Technology*  
Haifa, Israel  
ysteinbe@technion.ac.il

Michèle Wigger  
*LTCI Telecom Paris*  
91120 Palaiseau, France  
michele.wigger@telecom-paris.fr

*Abstract*—This article characterizes new boundary points on the capacity region of certain classes of more capable broadcast channels (BC) with uni-directional cooperation from the stronger to the weaker receiver. The new boundary points are achieved by a simple coding scheme that employs superposition coding at the transmitter with decode and forward at the stronger receiver. We evaluate our general result for Gaussian BCs and for a BC consisting of a binary erasure channel (BEC) to the stronger receiver and a binary symmetric channel (BSC) to the weaker receiver.

## I. Introduction

For non-cooperative broadcast channels (BC), the capacity region has been characterized for various classes of BCs. In particular, for BCs that are physically degraded (PD), stochastically degraded (SD), less noisy (LN), more capable (MC), essentially LN and MC [1], [2], and for the BCs with degraded message sets (asymmetric BCs, or ABCs, in the terminology of [3]). In all these BCs, superposition coding can achieve all rate-pairs in the capacity region.

For *cooperative* broadcast channels (CBC) [4]–[7], capacity results seem even more challenging. Nevertheless, the capacity region has been characterized for PD CBCs [5] and for one-sided cooperation in general ABCs [8], [9]. One sided-cooperation in semi-deterministic channels and cooperation for perfectly correlated Gaussian BCs were studied in [6]. Recently, a partial characterization of the capacity region of strongly less noisy channels with cooperation from the stronger user to the weaker was obtained in [7].

In most of these examples, capacity is exhausted by a simple scheme where the transmitter employs superposition coding (SPC) and the stronger (cooperative) user applies a decode and forward (D&F) [10] strategy so as to be able to send information to the weaker user that is only related to its desired message.

Our work is closest related to [7], which shows that above SPC and D&F strategy achieves a range of boundary points on the capacity region for certain classes of SD BCs with one-sided cooperation from the stronger to the weaker user and when the difference between the marginal channel strengths of the two users is large compared to the cooperation rate. In particular, [7] introduces a new *quantitative* notion of degradedness that applies to the CBC.

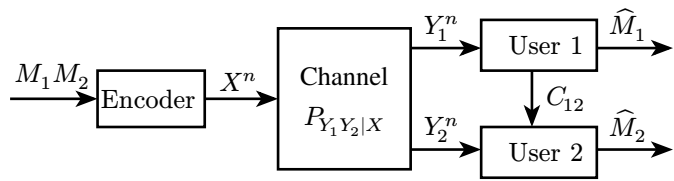

Fig. 1. Cooperative Broadcast Channel.

In this work, we improve on the results in [7]. That means, we consider MC CBCs with cooperation from the stronger user (User 1) to the weaker user (User 2), see Fig. 1, and we show that for some channels all rate-pairs $(R_1, R_2)$ in the cooperative capacity region with rate $R_1$ below a given threshold can be achieved using SPC and D&F. The same conclusion was already obtained in [7] for a larger set of CBCs but for a smaller threshold on the rate $R_1$ and under more stringent constraints on the cooperation rate $C_{12}$.

Specifically, for a BC with a binary erasure channel (BEC) to the stronger user and a binary symmetric channel (BSC) to the weaker user and for parameters so that the BC is MC, we provide a larger set of capacity boundary points where SPC and D&F is optimal compared to the results obtained from [7, Theorem 1]. The improved result in [7, Theorem 4] was not evaluated for this example, but leads to the same result as in this work. Direct evaluation of [7, Theorem 4] however seems more cumbersome than the proof in this paper.

The current work also presents boundary points for the Gaussian BC where SPC and D&F is optimal. The previous result in [7, Theorem 1] fails to provide optimality of this scheme for any boundary point on the Gaussian CBC. The improved result [7, Theorem 4] was again not evaluated for the Gaussian case.

We also present a general result for optimality of SPC and D&F, beyond Gaussian BCs and BCs consisting of a BEC and BSC. Our general result holds for all MC CBCs where the SPC and D&F inner bound and the outer bound provided in [7] have similar parametric expressions, the bounds differing only in an additional sum-rate constraint for the inner bound. In our proof, we follow the idea in [7] and establish optimality of the SPC and D&F inner bound whenever the sum-rate constraint is not active. Unlike [7], which employs a statistical approach to show that the sum-rate constraint is not active, in this work, we directly exploit the parametric characterisation.

The manuscript is organized as follows. The problem setup is introduced in Section II and the main results are presented in Section III. Proofs are provided in Section IV.

## II. PROBLEM SETUP

*A. Channel Model and Preliminaries*

A two-receiver broadcast channel (BC) $P_{Y_1,Y_2|X}$ consists of one input over a given alphabet $\mathcal{X}$ and two outputs over respective alphabets $\mathcal{Y}_1$ and $\mathcal{Y}_2$. Under uni-directional cooperation, a link from User 1 to User 2, with capacity $C_{12}$, allows the users to cooperate.

For a given blocklength $n$, the encoder picks for each user $k = 1, 2$ a uniform message $M_k$ over the set $\mathcal{N}_k = \{1, ..., \nu_k\}$. User 1 picks a cooperation message $L_c$ from the set $\mathcal{N}_c = \{1, ..., 2^{nC_{12}}\}$ and in function of its observed outputs $Y_1^n$. Moreover, each of the two users $k = 1, 2$ attempts to decode its intended message $M_k$ by producing a guess $\widehat{M}_k$. User 1 produces its guess based on its observed outputs $Y_1^n$ only, and User 2 based on both its observed outputs $Y_2^n$ and the cooperation message $L_c$.

Accordingly, we have the following definitions:

**Definition 1.** An $(n, \nu_1, \nu_2, 2^{nC_{12}}, \varepsilon)$ code for a cooperative BC consist of a source encoder

$$f : \mathcal{N}_1 \times \mathcal{N}_2 \to \mathcal{X}^n, \tag{1}$$

a cooperation encoder

$$f_c : \mathcal{Y}_1^n \to \mathcal{N}_c, \tag{2}$$

two decoders

$$\varphi_1 : \mathcal{Y}_1^n \to \mathcal{N}_1, \tag{3.1}$$
$$\varphi_2 : \mathcal{Y}_2^n \times \mathcal{N}_c \to \mathcal{N}_2, \tag{3.2}$$

and a threshold on the probability of error

$$\frac{1}{\nu_1 \nu_2} \sum_{(j_1,j_2) \in \mathcal{N}_1 \times \mathcal{N}_2} P_{Y_1 Y_2|X}^{\otimes n}\left(D_{j_1 j_2}^c \mid f(j_1, j_2)\right) \le \varepsilon, \tag{4}$$

where $D_{j_1 j_2}$ is the decision region of the messages $(j_1, j_2)$ and $D_{j_1 j_2}^c$ its complement.

**Definition 2.** A pair of rates $(R_1, R_2)$ is called achievable with cooperation capacity $C_{12}$ if for any given $\delta > 0$, $\varepsilon > 0$ and sufficiently large blocklengths $n$, there exists an $(n, 2^{n(R_1-\delta)}, 2^{n(R_2-\delta)}, 2^{nC_{12}}, \varepsilon)$ code.

The capacity region, denoted $\mathcal{C}(C_{12})$, is the closure of the set of all achievable pairs $(R_1, R_2)$ for a given cooperation capacity $C_{12} \ge 0$.

For positive cooperation capacity $C_{12} > 0$, the capacity region $\mathcal{C}(C_{12})$ depends on the joint conditional law $P_{Y_1,Y_2|X}$, while without cooperation it depends only on the marginal transition probabilities $P_{Y_1|X}$ and $P_{Y_2|X}$.

A prominent and simple coding strategy is to use superposition coding (SPC) at the transmitter and decode-and-forward (D&F) at User 1. In this context, D&F relies on binning: the set $\mathcal{N}_2$ is divided into $2^{n\mathcal{N}_c}$ bins, each containing $2^{n[R_2-C_{12}]_+}$ elements. User 1 performs joint decoding of $M_1$ and $M_2$ and chooses the cooperation message as the bin number of its estimated $M_2$. As described in [7, Appendix B], this strategy achieves all rate pairs in the region $\mathcal{R}_{in}(C_{12})$, defined as the set of rate pairs $(R_1, R_2)$ satisfying

$$R_1 \le I(X; Y_1|U), \tag{5.1}$$
$$R_2 \le I(U; Y_2) + C_{12}, \tag{5.2}$$
$$R_1 + R_2 \le I(X; Y_1), \tag{5.3}$$

for some joint distribution $P_{UX} P_{Y_1 Y_2|X}$.

The region $\mathcal{R}_{in}(C_{12})$ is known to coincide with the capacity region only in the special cases of PD CBCs [5] or less noisy CBCs when $C_{12} = 0$, i.e., without cooperation. It also achieves capacity for arbitrary ABCs [8], [9].

A valid outer bound on capacity for more capable (MC) BCs [7, Eq. 69] is $\mathcal{R}_{out}(C_{12})$, defined as the pairs $(R_1, R_2)$ satisfying

$$R_1 \le I(X; Y_1|U), \tag{6.1}$$
$$R_2 \le I(U; Y_2) + C_{12}, \tag{6.2}$$

for some joint distribution $P_{UX} P_{Y_1 Y_2|X}$.

In the following throughout we assume that the BC $P_{Y_1 Y_2|X}$ is MC, i.e.,

$$I(X; Y_1) \ge I(X; Y_2), \forall P_X, \tag{7}$$

and can thus write [7]:

$$\mathcal{R}_{in}(C_{12}) \subseteq \mathcal{C}(C_{12}) \subseteq \mathcal{R}_{out}(C_{12}). \tag{8}$$

Observe that both $\mathcal{R}_{in}(C_{12})$ and $\mathcal{R}_{out}(C_{12})$ depend only on $P_{Y_1|X}$ and $P_{Y_2|X}$ while $\mathcal{C}(C_{12})$ depends on the $P_{Y_1,Y_2|X}$.

Note that the bounds (5) and (6) differ only in the additional sum-rate constraint (5.3). This fact was exploited in [7] to show that if for some range of rates $(C_{12}, R_1, R_2)$, (5.1) and (5.2) imply (5.3), inner and outer bounds coincide, yielding a capacity characterization for that range. The main contribution in [7] is to find statistical conditions (so called strong LN conditions) under which the sum-rate bound (5.3) is implied by the individual rate bounds. In the present work, we exploit parametric characterizations of inner and outer bounds to identify such ranges of rates $(C_{12}, R_1, R_2)$.

## III. RESULTS

Let $C_1$ and $C_2$ denote the marginal capacities of the channels $P_{Y_1|X}$ and $P_{Y_2|X}$. For any $R_1 \le C_1$, let $R_2^*(R_1)$ denote the largest rate $R_2$ so that $(R_1, R_2)$ are achievable:

$$R_2^*(R_1) = \sup\{R_2 : (R_1, R_2) \in \mathcal{C}(C_{12})\}. \tag{9}$$

We start by stating a general result, which we then specialize to Gaussian BCs and to BCs that consist of a BEC to the stronger User 1 and a BSC to the weaker User 2.

*A. General Result*

**Proposition 1:** Consider a channel $P_{Y_1 Y_2|X}$ satisfying the following condition. The regions $\mathcal{R}_{in}(C_{12})$ and $\mathcal{R}_{out}(C_{12})$ can both be expressed in parametric form of a common parameter:

$$\mathcal{R}_{in}(C_{12}) = \bigcup_{\alpha \in [0,b]} \mathcal{R}_{in}^\alpha(C_{12}) \tag{10}$$

and

$$\mathcal{R}_{out}(C_{12}) = \bigcup_{\alpha \in [0,b]} \mathcal{R}_{out}^\alpha(C_{12}) \tag{11}$$

where

$$\mathcal{R}_{out}^\alpha(C_{12}) = \{(R_1, R_2) : R_1 \leq f_1(\alpha) \quad (12.1)$$
$$R_2 \leq f_2(\alpha) \quad (12.2)$$
$$R_1 + R_2 \leq C_1 \quad\}, \quad (12.3)$$

and

$$\mathcal{R}_{out}^\alpha(C_{12}) = \{(R_1, R_2) : R_1 \leq f_1(\alpha) \quad (13.1)$$
$$R_2 \leq f_2(\alpha)\}, \quad (13.2)$$

where

- $f_1$ is a continuous and strictly *increasing* function with $f_1(0) = 0$ and $f_1(b) = C_1$;
- $f_2$ is a continuous and strictly *decreasing* function with $f_2(0) = C_2 + C_{12}$ and $f_2(b) = C_{12}$;
- the sum $f_1 + f_2$ is strictly increasing.

Assume that

$$C_{12} \leq C_1 - C_2, \quad (14)$$

and define $\alpha_{th} \leq b$ as the unique solution in $[0, b]$ to

$$f_1(\alpha_{th}) + f_2(\alpha_{th}) = C_1 \quad (15)$$

and

$$R_{1,th} := f_1(\alpha_{th}). \quad (16)$$

Notice that (15) has the desired unique solution because the left-hand side of (15) is continuous and strictly increasing, and for $\alpha_{th} = 0$ it evaluates to $C_2 + C_{12}$, which by Assumption (14) is smaller than $C_1$, and for $\alpha_{th} = b$ it evaluates $C_1 + C_{12}$ which is larger than $C_1$.

Then, all pairs $(R_1, R_2)$ in the capacity region $\mathcal{C}(C_{12})$ with

$$R_1 \leq R_{1,th} \quad (17)$$

are achieved using SPC and D&F at User 1. Thus, the inner bound $\mathcal{R}_{in}(C_{12})$ is tight in this regime.

In particular, for $R_1 \leq R_{1,th}$:

$$R_2^*(R_1) = f_2(f_1^{-1}(R_1)) \quad (18)$$

and the maximum sum-rate is bounded as:

$$R_1 + R_2^*(R_1) \leq C_1, \quad (19)$$

where the inequality is strict unless $R_1 = R_{1,th}$.

**Remark 1.** The thresholds $\alpha_{th}$ and $R_{1,th}$ are continuous and decreasing functions of $C_{12}$ ranging from $b$ and $C_1$, respectively, when $C_{12} = 0$, to 0 when $C_{12} = C_1 - C_2$.

Above proposition and remark are proved in Section IV.A.

In the following, we specialize above proposition to the Gaussian BC where User 1 has larger capacity than User 2, as well as to BCs consisting of a BEC to User 1 and a BSC to User 2 for all parameter ranges so that User 1 is MC than User 2. Notice that Proposition 1 provides also similar results for MC BCs consisting of two BSCs or two BECs. Details for these scenarios are omitted due to lack of space.

### B. The Gaussian BC

Consider the Gaussian BC

$$Y_k = \sqrt{s_k}X + Z_k, \quad k = 1, 2, \quad (20)$$

where $Z_1$ and $Z_2$ are jointly Gaussian of unit variances and $s_1$ and $s_2$ are given constants $s_1 > s_2 > 0$.

The capacities of the two marginal channels are $C_k = \Gamma(s_k)$, for $k = 1, 2$, where we define the Gaussian capacity function $\Gamma(x) = \log(1 + x)/2$. Note that the function $\Gamma$ is invertible.

Define further, for $C_{12} \leq C_1 - C_2$,

$$\alpha_{th} := \left( \frac{s_1 - s_2}{\Gamma^{-1}(C_1 - C_2 - C_{12})} - s_2 \right)^{-1} \quad (21)$$

and

$$R_{1,th} := \Gamma(\alpha_{th} s_1). \quad (22)$$

**Theorem 1.** Consider above Gaussian cooperative BC and assume that

$$C_{12} \leq C_1 - C_2. \quad (23)$$

All rate-pairs $(R_1, R_2)$ in the capacity region $\mathcal{C}(C_{12})$ with

$$R_1 \leq R_{1,th}, \quad (24)$$

are achieved using SPC and D&F at User 1.

Furthermore, for $R_1 \leq R_{1,th}$:

$$R_2^*(R_1) = C_2 + C_{12} - \Gamma\left(\Gamma^{-1}(R_1)\frac{s_2}{s_1}\right) \quad (25)$$

and the sum-rate is bounded by $C_1$:

$$R_1 + R_2^*(R_1) \leq C_1, \quad (26)$$

where the inequality is strict unless $R_1 = R_{1,th}$.

Above theorem follows from Proposition 1. Details are provided in Section IV.B.

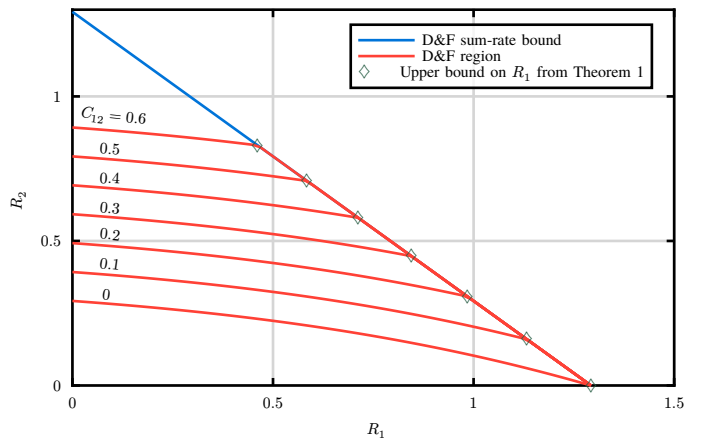

Fig. 2. Region $\mathcal{R}_{in}(C_{12})$ for $s_1 = 5, s_2 = 0.5$, and different values of $C_{12}$.

Fig. 2 depicts the region $\mathcal{R}_{in}(C_{12})$ for this Gaussian example when $s_1 = 5$, $s_2 = 0.5$, and different values of the cooperation capacity $C_{12}$. For positive values of $C_{12} > 0$, our theorem shows that it is optimal for all boundary points with sum-rate $R_1 + R_2 < C_1$ and the left-most point with $R_1 + R_2 = C_1$. This limiting boundary point is indicated with a diamond symbol in the figure. For all boundary points of $\mathcal{R}_{in}(C_{12})$ to the right, the sum-rate equals $C_1$. We conjecture that these points lie in the interior of the capacity region, and not on its boundary, and thus SPC and D&F at User 1 are optimal if, and only if, $R_1 + R_2 \leq C_1$.

### C. A BC with a BEC to User 1 and a BSC to User 2

We next consider a $\text{BEC}(\tau_1)$ to User 1 and a $\text{BSC}(p_2)$ to User 2, where we assume $p_2 \in [0, 1/2]$ and moreover
$$0 \leq \tau < H_b(p_2). \tag{27}$$
These conditions ensure that the BEC is MC than the BSC [1, example 5.4].

For $C_{12} \leq C_1 - C_2$, define now $q_{th}$ as the unique solution in $[0, 1/2]$ to the equation
$$H_b(p_2 \star q_{th}) - H_b(q_{th})(1 - \tau_1) = C_{12} + \tau_1 \tag{28}$$
where $H_b$ denotes the binary entropy function. Define further
$$R_{1,th,2} := H_b(q_{th})(1 - \tau_1). \tag{29}$$

**Theorem 2.** Consider above CBC formed by a $\text{BEC}(\tau_1)$ to User 1 and a $\text{BSC}(p_2)$ to User 2 and assume that $C_{12} \leq C_1 - C_2$.

All rate-pairs $(R_1, R_2)$ in the capacity region $\mathcal{C}(C_{12})$ with
$$R_1 \leq R_{1,th,2} \tag{30}$$
are achieved using SPC and D&F at User 1.

Furthermore, for $R_1 \leq R_{1,th,2}$:
$$R_2^*(R_1) = 1 - H_b\left(p_2 \star H_b^{-1}\left(\frac{R_1}{1 - \tau_1}\right)\right) + C_{12}, \tag{31}$$
where $H_b^{-1}$ denotes the inverse of the binary entropy function on the domain $[0, 1/2]$.

For all $R_1 \leq R_{1,th,3}$, the sum-rate is bounded by $C_1$:
$$R_2^*(R_1) + R_1 \leq C_1, \tag{32}$$
where the inequality is strict except for $R_1 = R_{1,th,2}$.

The proof is again obtained from Proposition 1 and details are given in Section IV.C.

The following Fig. 3 shows the region $\mathcal{R}_{in}(C_{12})$ for a BC consisting of a BEC(0.1) to User 1 and a BSC(0.2) to User 2, and different values of the cooperation capacity $C_{12}$.

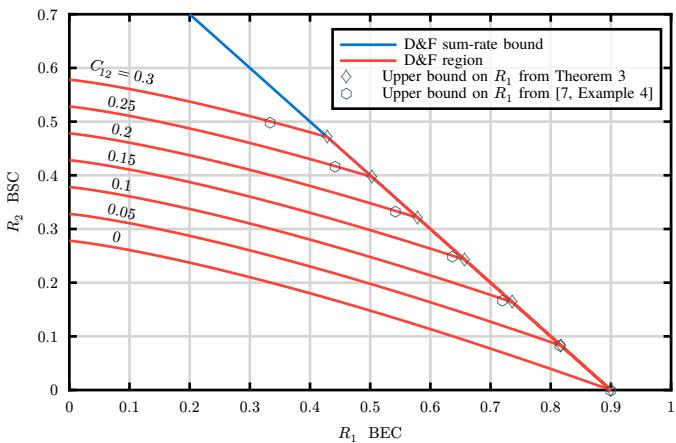

Fig. 3. Region $\mathcal{R}_{in}(C_{12})$ for the BEC(0.1)-BSC(0.2) BC, and different values of $C_{12}$.

Our theorem shows that $\mathcal{R}_{in}(C_{12})$ coincides with the capacity region for all boundary points with sum-rate $R_1 + R_2 < C_1$ as well as the left-most boundary point where $R_1 + R_2 = C_1$. This limiting boundary point is indicated with a diamond symbol in the figure. The hexagon symbols indicate the limiting points were the previous work [7] showed optimality of $\mathcal{R}_{in}(C_{12})$ (and thus SPC and D&F).

From the figure we see that the current work establishes the optimality of SPC and D&F on a larger part of the capacity region, and extends the class of channels and cooperation rates $C_{12}$ for which D&F is optimal, compared to the results in [7, Theorem 1].

## IV. Proofs

### A. Proof of Proposition 1

Define for each $R_1 \leq C_1$:
$$\bar{R}_2(R_1) := \sup\{R_2 : (R_1, R_2) \in \mathcal{R}_{out}(C_{12})\}. \tag{33}$$
We start by proving that
$$\mathcal{R}_{in}^\alpha(C_{12}) = \mathcal{R}_{out}^\alpha(C_{12}), \forall \alpha \leq \alpha_{th}. \tag{34}$$
To this end, notice that $\mathcal{R}_{out}^\alpha(C_{12})$ is a rectangular region, and $\mathcal{R}_{in}^\alpha(C_{12})$ coincides with this rectangular region whenever its sum-rate constraint is not active:
$$f_1(\alpha) + f_2(\alpha) \leq C_1. \tag{35}$$
Since the sum $f_1 + f_2$ is a continuous and strictly increasing function, this inequality is satisfied for all $\alpha \leq \alpha_{th}$, establishing the desired Equality (34).

Moreover, since $f_1 + f_2$ is strictly increasing, it can be deduced that for all $\alpha \leq \alpha_{th}$ it holds that $f_1(\alpha) + f_2(\alpha) \leq C_1$, where the inequality is strict unless $\alpha = \alpha_{th}$. This implies that the sum-rate of the dominant corner point (and thus of all points) of $\mathcal{R}_{out}^\alpha(C_{12})$ is bounded by $C_1$, where equality only holds when $\alpha = \alpha_{th}$.

We next argue that each boundary point $(R_1, \bar{R}_2(R_1))$ of $\mathcal{R}_{out}(C_{12})$ with $R_1 \leq R_{1,th}$ is the dominant corner point of the rectangle $\mathcal{R}_{out}^{\alpha(R_1)}(C_{12})$ for
$$\alpha(R_1) := f_1^{-1}(R_1). \tag{36}$$
Since $\alpha(R_{1,th}) = \alpha_{th}$, and $f_1^{-1}$ is continuous and strictly increasing, we can then deduce by (34) that $(R_1, \bar{R}_2(R_1))$ also lies in $\mathcal{R}_{in}^{\alpha(R_1)}(C_{12})$ and is thus achievable using SPC and D&F. To see that $\mathcal{R}_{out}(C_{12})$ with $R_1 \leq R_{1,th}$ is the dominant corner point of the rectangle $\mathcal{R}_{out}^{\alpha(R_1)}(C_{12})$, it suffices to notice that the largest $R_1$ in $\mathcal{R}_{out}^\alpha(C_{12})$ is continuous and strictly increasing in $\alpha$ and the largest $R_2$ continuous and strictly decreasing in $\alpha$. This establishes the first part of the proposition.

We continue to notice that from the considerations around (35), we can conclude that for all $R_1 \leq R_{1,th}$ we have
$$R_1 + \bar{R}_2(R_1) \leq C_1, \tag{37}$$
because $\alpha(R_1) \leq \alpha_{th}$. And moreover, the inequality is strict unless $\alpha(R_1) = \alpha_{th}$ and thus $R_1 = R_{1,th}$.

To establish the last two desired statements (18) and (19) of the proposition, it suffices to notice that $(R_1, \bar{R}_2(R_1))$ coincides with $(R_1, R_2^*(R_1))$ because $\mathcal{R}_{out}^\alpha(C_{12})$ lies on the boundary of an outer bound. We therefore, have by the definition of $\mathcal{R}_{out}^\alpha(C_{12})$,
$$R_2^*(R_1) = f_2(f_1^{-1}(R_1)). \tag{38}$$
and by (37),
$$R_1 + R_2^*(R_1) \leq C_1. \tag{39}$$

### B. Proof of Theorem 1

We verify that the conditions in Proposition 1 are satisfied.

Using standard arguments and the Entropy-Power Inequality (EPI), similarly to the non-cooperative case, it can be shown that jointly Gaussian $P_{UX}$ exhaust the regions $\mathcal{R}_{in}(C_{12})$ and $\mathcal{R}_{out}(C_{12})$ [1, Chap. 5.5]. Therefore, the following parametric forms hold:
$$\mathcal{R}_{in}(C_{12}) = \bigcup_{\alpha \in [0,1]} \mathcal{R}_{in}^\alpha(C_{12}), \tag{40}$$
and
$$\mathcal{R}_{out}(C_{12}) = \bigcup_{\alpha \in [0,1]} \mathcal{R}_{out}^\alpha(C_{12}) \tag{41}$$
for
$$\mathcal{R}_{in}^\alpha(C_{12}) := \{(R_1, R_2) : R_1 \leq \Gamma(\alpha s_1), \tag{42.1}$$
$$R_2 \leq C_2 + C_{12} - \Gamma(\alpha s_2), \tag{42.2}$$
$$R_1 + R_2 \leq C_1\}, \tag{42.3}$$
and
$$\mathcal{R}_{out}^\alpha(C_{12}) := \{(R_1, R_2) : R_1 \leq \Gamma(\alpha s_1), \tag{43.1}$$
$$R_2 \leq C_2 + C_{12} - \Gamma(\alpha s_2)\}. \tag{43.2}$$
Notice that the function $f_1(\alpha) = \Gamma(\alpha s_1)$ is continuous, strictly increasing in $\alpha$, and satisfies $f_1(0) = 0$ and $f_1(1) = C_1$. Similarly, the function $f_2(\alpha) = C_2 + C_{12} - \Gamma(\alpha s_2)$ is continuous, strictly decreasing in $\alpha$, and satisfies $f_2(0) = C_2 + C_{12}$ and $f_2(1) = C_{12}$. Moreover, the sum
$$f_1(\alpha) + f_2(\alpha) = \frac{1}{2}\log\left(\frac{1+\alpha s_1}{1+\alpha s_2}\right) \tag{44.1}$$
$$= C_2 + C_{12} + \frac{1}{2}\log\left(1 + \frac{s_1 - s_1}{\frac{1}{\alpha} + s_2}\right) \tag{44.2}$$
is strictly increasing in $\alpha$.

We thus conclude that above parametric form satisfies all the conditions in Proposition 1.

It remains to verify that the proposed values for $\alpha_{th}$ and $R_{1,th}$ as well as for $R_2^*(R_1)$ are as in Proposition 1. By Proposition 1, $\alpha_{th}$ is the unique solution to
$$\Gamma(\alpha_{th} s_1) + C_2 + C_{12} - \Gamma(\alpha_{th} s_2) = C_1, \tag{45}$$
which is equivalent to
$$C_1 - C_2 - C_{12} = \frac{1}{2}\log\left(1 + \frac{s_1 - s_2}{\frac{1}{\alpha_{th}} + s_2}\right) \tag{46}$$
and can be rewritten as
$$\alpha_{th} = \left(\frac{s_1 - s_2}{\Gamma^{-1}(C_1 - C_2 - C_{12})} - s_2\right)^{-1}, \tag{47}$$
which corresponds to the definition in Theorem 1. Moreover, $R_{1,th} = f_1(\alpha_{th}) = \Gamma(\alpha_{th} s_1)$ as indicated in the theorem.

Finally, by Proposition 1,
$$R_2^*(R_1) = f_2(f_1^{-1}(R_1)) \tag{48.1}$$
$$= C_2 + C_{12} - \Gamma\left(\Gamma^{-1}(R_1)\frac{s_2}{s_1}\right). \tag{48.2}$$
This concludes the proof of Theorem 1.

### C. Proof of Theorem 2

We verify that the conditions in Proposition 1 are satisfied.

Using standard arguments, involving Mrs. Gerber's Lemma [11], it is shown in Appendix A that
$$\mathcal{R}_{in}(C_{12}) = \bigcup_{q \in [0,1/2]} \mathcal{R}_{in}^q(C_{12}), \tag{49}$$
and
$$\mathcal{R}_{out}(C_{12}) = \bigcup_{q \in [0,1/2]} \mathcal{R}_{out}^q(C_{12}). \tag{50}$$
where for given $q \in [0, 1/2]$:
$$\mathcal{R}_{in}^q(C_{12}) := \{(R_1, R_2) : R_1 \leq \mathrm{H_b}(q)(1-\tau_1), \tag{51.1}$$
$$R_2 \leq 1 - \mathrm{H_b}(p_2 \star q) + C_{12}, \tag{51.2}$$
$$R_1 + R_2 \leq 1 - \tau_1\}, \tag{51.3}$$
and
$$\mathcal{R}_{out}^q(C_{12}) := \{(R_1, R_2) : R_1 \leq \mathrm{H_b}(q)(1-\tau_1), \tag{52.1}$$
$$R_2 \leq 1 - \mathrm{H_b}(p_2 \star q) + C_{12}\}, \tag{52.2}$$
Notice that the function
$$f_1(q) = \mathrm{H_b}(q)(1-\tau_1) \tag{53}$$
is continuous and strictly increasing in $q$, with $f_1(0) = 0$ and $f_1(1/2) = C_1$. Similarly, the function
$$f_2(q) = 1 - \mathrm{H_b}(p_2 \star q) + C_{12} \tag{54}$$
is continuous and strictly decreasing in $q$, with $f_2(0) = C_2 + C_{12}$ and $f_2(1/2) = C_{12}$. Moreover, the sum
$$f_1(q) + f_2(q) = \mathrm{H_b}(q)(1-\tau_1) + 1 - \mathrm{H_b}(p_2 \star q) + C_{12} \tag{55}$$
is strictly increasing in $q$ as can be seen by examining the derivative.

We thus conclude that above parametric form satisfies the conditions in Proposition 1. It remains to verify that the values for $q_{th}$ and $R_{1,th,2}$, as well as, for $R_2^*(R_1)$ proposed in the theorem are as in Proposition 1.

By Proposition 1, $q_{th}$ is the unique solution to
$$\mathrm{H_b}(q)(1-\tau_1) + 1 - \mathrm{H_b}(p_2 \star q) + C_{12} = C_1 \tag{56}$$
which is equivalent to the expression proposed in the theorem. Furthermore, $R_{1,th} = f_1(q_{th})$, also as proposed in the theorem. Finally, by Proposition 1:
$$R^*(R_1) = f_2(f_1^{-1}(R_1)) \tag{57.1}$$
$$= 1 - \mathrm{H_b}\left(p_2 \star \mathrm{H_b}^{-1}\left(\frac{R_1}{1-\tau_1}\right)\right) + C_{12}. \tag{57.2}$$
This concludes the proof.

ACKNOWLEDGMENT

The authors acknowledge funding support from the ERC under Grant Agreement 101125691.

## APPENDIX

*A. Proofs of (49) and (50)*

The inclusion $\supseteq$ is obtained by restricting the joint law $P_{UX}$ in (5) and (6) to $U$ Bernoulli-$\frac{1}{2}$ and $P_{X|U}$ is a BSC($q$) for some $q \in [0, 1/2]$.

To prove the inclusion $\subseteq$, notice

$$\mathrm{I}(X;Y_1|U) = \mathrm{I}(X;Y_1) - \mathrm{I}(U;Y_1) \qquad (58.1)$$
$$= \tau_1 \mathrm{H}(X) - \tau_1 \mathrm{I}(U;X) \qquad (58.2)$$
$$= \tau_1 \mathrm{H}(X|U). \qquad (58.3)$$

On the other hand,

$$\mathrm{I}(U;Y_2) \leq 1 - \mathrm{H}(Y_2|U), \qquad (59)$$

and, since $\mathrm{H}(Y_2|X) \leq \mathrm{H}(Y_2|U) \leq 1$, there exists a $q \in [0, 1/2]$ such that

$$\mathrm{H}(Y_2|U) = \mathrm{H}_{\mathrm{b}}(p_2 \star q). \qquad (60)$$

Then, due to Mrs. Gerber's Lemma [11],

$$\mathrm{H}_{\mathrm{b}}\big(\mathrm{H}_{\mathrm{b}}^{-1}(\mathrm{H}(X|U)) \star p_2\big) \leq \mathrm{H}(Y_2|U), \qquad (61)$$

which gives

$$\mathrm{H}(X|U) \leq \mathrm{H}_{\mathrm{b}}(q). \qquad (62)$$

We thus conclude that $\mathcal{R}_{out}(C_{12})$ is contained in the set of all rate-pairs $(R_1, R_2)$ satisfying

$$R_1 \leq \mathrm{H}_{\mathrm{b}}(q)(1 - \tau_1), \qquad (63.1)$$
$$R_2 \leq 1 - \mathrm{H}_{\mathrm{b}}(p_2 \star q) + C_{12}. \qquad (63.2)$$

Similarly, $\mathcal{R}_{in}(C_{12})$ is contained in the set of all rate-pairs $(R_1, R_2)$ satisfying

$$R_1 \leq \mathrm{H}_{\mathrm{b}}(q)(1 - \tau_1), \qquad (64.1)$$
$$R_2 \leq 1 - \mathrm{H}_{\mathrm{b}}(p_2 \star q) + C_{12}, \qquad (64.2)$$
$$R_1 + R_2 \leq C_1. \qquad (64.3)$$